# Full-volume three-component intraventricular vector flow mapping by triplane color Doppler

F. Vixège, A. Berod, P.Y. Courand, S. Mendez, F. Nicoud, P. Blanc-Benon, D. Vray and D. Garcia

*Abstract* — **Intraventricular vector flow mapping (iVFM) is a technique for retrieving 2-D velocity vector fields of blood flow in the left ventricle. This method is based on conventional color Doppler imaging, which makes iVFM compatible with the clinical setting. We have generalized the iVFM for a three-dimensional reconstruction (3D-iVFM). 3D-iVFM is able to recover three-component velocity vector fields in a full intraventricular volume by using a clinical echocardiographic triplane mode. As with the 2-D version, the method is based on the mass conservation, and free-slip boundary conditions on the endocardial wall. These mechanical constraints were imposed in a least-squares minimization problem that was solved through the method of Lagrange multipliers. We validated 3D-iVFM *in silico* in a patient-specific CFD (computational fluid dynamics) model of cardiac flow, and tested its feasibility *in vivo* on volunteers. In both *in silico* and *in vivo* investigations, the dynamics of the intraventricular vortex that forms during diastole was deciphered by 3D-iVFM. Our results tend to indicate that 3D-iVFM could provide full-volume echocardiographic information on left intraventricular hemodynamics from the clinical modality of triplane color Doppler.**

## INTRODUCTION

DURING diastole, blood from the left atrium flowing through the mitral valve forms a vortex in the left ventricle. This ring-shaped vortex expands and stretches, causing blood to swirl in a natural direction that helps redirect flow circulation to the outflow tract, while limiting energy loss [1], [2]. When filling is impaired (i.e., in the presence of diastolic dysfunction), there is a change in cardiac dynamics that significantly affects the shape and properties of the vortex. Recent works addressing the intraventricular flow suggest that the quantification of the vortex could improve the diagnosis of diastolic dysfunction. Echocardiographic analysis of the vortex is then an avenue to consider [3]–[5], given that the standard clinical indices of diastolic function often lead to divergent conclusions in patients.

The two most common imaging techniques used to study intracardiac blood flow are CMR (cardiac magnetic resonance) and Doppler echocardiography. CMR allows the clinician to collect 4-D (3-D + time) information of the flow, from acquisitions on a few consecutive cardiac cycles. Its operating time, from 30 min to 60 min per patient, as well as the examination costs, make it a solution primarily intended for clinical research [6], [7]. In contrast, Doppler echocardiography is the most widely used clinical modality for blood flow analyses. Doppler echo, in its most classical version, yields only one component of the velocity vectors, the one parallel to the direction of the ultrasound wave. To overcome this limitation, several ultrasound imaging techniques have been developed for a 2-D vector depiction of the intracardiac flow, such as echo-PIV (echographic particle image velocimetry) by contrast-enhanced echocardiography [8], [9], blood speckle tracking [10], or iVFM (intraventricular vector flow mapping) [11], [12]. The echo-PIV technique requires the intravenous injection of microbubbles, which is not compatible with a routine clinical use. Blood speckle tracking has the advantage of being free of microbubbles and uses ultrasound signals backscattered by red blood cells [13], [14]. Sensitive to the clutter signals, this technique nevertheless seems well adapted to pediatric echocardiography where spatial resolution is higher [15].

To get a broader view of the heart circulation and to overcome the clinical limitations of 3-D CMR velocimetry, some research groups have also focused on 3-D reconstruction of intracardiac flows by ultrasound. Some of the methods used were derived from 2-D techniques. Feasibility of echo-PIV to obtain a volumetric view of intraventricular flow has been described *in vitro* and *in vivo* [16], [17], [9]. Gomez *et al.* recovered the 3-D flow [18], [19] by spatio-temporally registering three or four color Doppler volumes acquired from different acoustic windows. These two approaches cannot be used routinely for clinical purposes because of their methodological constraints (microbubbles, multi-view). Inspired by 2-D iVFM, Grønli *et al.* reconstructed 3-D intraventricular flow from Doppler volumes with an optimization method that aimed to conserve the mass (continuity equation) and momentum (Navier-Stokes equations). In this study, the intracardiac pressure gradients were omitted in the Navier-Stokes equations to simplify the numerical problem. After writing the velocity fields with B-splines, the problem

This work was supported by the LABEX CeLyA (ANR-10-LABX-0060) of Université de Lyon, within the program « Investissements d'Avenir » (ANR-16IDEX-0005) op-erated by the French National Research Agency (ANR). F. Vixege, D. Vray and D. Garcia are with the Univ Lyon, INSA-Lyon, CREATIS, UMR5220, U1294, France. (e-mail: florian.vixege@creatis.insa-lyon.fr; damien.garcia@inserm.fr; garcia.damien@gmail.com). A. Berod, S. Mendez and F. Nicoud are with the University of Montpellier—IMAG CNRS UMR 5149, France. P.Y. Courand is with the Department of Cardiology, Hospices Civils de Lyon, France. P. Blanc-Benon is with the Univ Lyon, École Centrale de Lyon, LMFA UMR 5509, France.

was solved using the open-source package TensorFlow [20]. Although closer to the clinical situation, it remains difficult to rely on Doppler volumes because of their low temporal resolution.

To obtain a three-dimensional outlook of the left intraventricular flow, we hypothesized that our 2-D iVFM algorithm [12], [21] could be generalized to 3-D by using the triplane mode available on clinical ultrasound scanners. In triplane echocardiography, 2-, 3-, and 4-chamber views are displayed simultaneously with 60-degree interplane angles (Figure 1). The triplane mode has shown clinical relevace in echocardiography laboratories. For example, its diagnostic advantages have been evaluated in the calculation of ejection fraction [22], [23] and global longitudinal strain by speckle-tracking [24]. From these studies, it appears that these three planes contain sufficient information for an accurate analysis of the global volume and deformations of the left ventricle. In the same vein, we postulated that the Doppler velocities of these three planes also contain sufficient information for an adequate construction of the three-dimensional intraventricular flow. This would be an advantage over volume Doppler because triplane color Doppler provides a higher imaging rate.

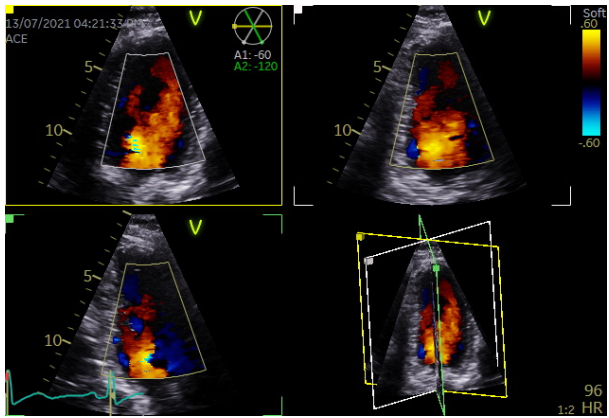

Figure 1 – Triplane color Doppler during a routine echocardiographic examination. Two-, three-, and four-chamber views are displayed simultaneously with 60-degree interplane angles.

In this paper, we introduce a 3-D iVFM technique based on triplane Doppler-echo, which generalizes the 2-D iVFM developed by Assi *et al.* [12]. The 3D-iVFM allows the echographer to estimate the 3-D intraventricular flow from three color-Doppler planes, by using a minimization problem constrained by hemodynamic properties (mass conservation in the cardiac cavity, and free-slip boundary conditions on the endocardium). The physical equality constraints were imposed using the Lagrange multiplier method. The regularization parameter that controls the smoothing was determined automatically through the *L*-curve method. In the following, we describe the 3D-iVFM method and its numerical formulation, its validation using a patient-specific CFD (computational fluid dynamics) model, and its clinical feasibility through *in vivo* examples.

## Method

### Triplane Mode

The objective of 3D-*i*VFM (3D-*intraventricular* Vector Flow Mapping) is to recover full-volume three-component velocity fields, in the left ventricular cavity, by using triplane color Doppler. Figure 1 illustrates a triplane color Doppler, which provides Doppler velocities from three different apical long-axis views separated by a 60° angle: the two-chamber, three-chamber and four-chamber views. We chose the triplane mode (rather than volume Doppler) because 1) the ultrasound data before scan-conversion are available (Figure 2) through the workstation EchoPAC (GE Healthcare), and 2) the temporal resolution is higher. The EchoPAC software gives access to the Doppler velocities and 8-bit B-mode images in the spherical coordinate system $\{r, \theta, \varphi\}$ related to the cardiac phased array (Figure 3). In this spherical system, $r \geq 0$, $\theta \in [0, \pi/2]$, and $\varphi \in [0, 2\pi]$.

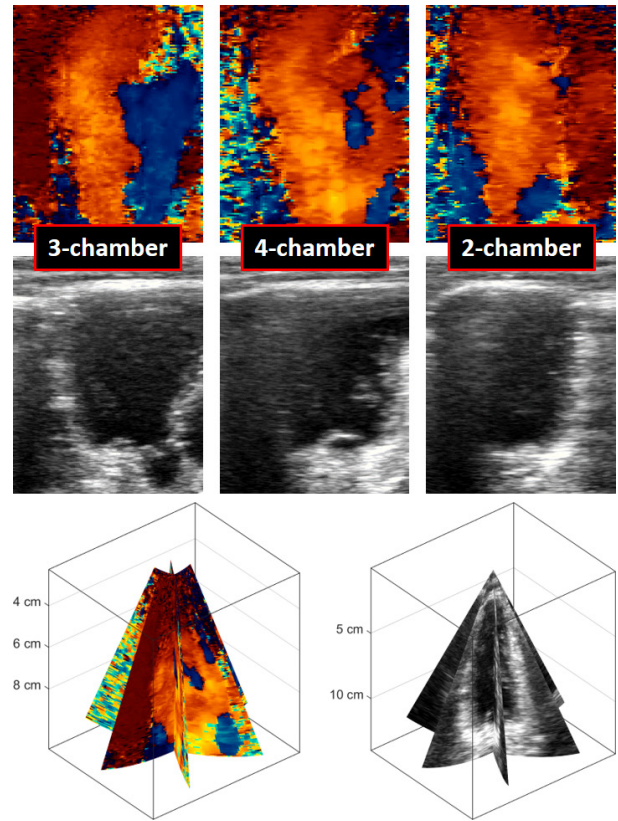

Figure 2 – In the 3D-iVFM (intraventricular Vector Flow Mapping) technique, we use the Doppler velocities (1st row) and endocardial displacements (from the B-mode images, 2nd row) before scan-conversion, i.e. in the coordinate system defined in Figure 3. The 3rd row is the corresponding triplane view.

### iVFM as a Constrained Minimization Problem

In this coordinate system, the Doppler velocities ($u_\mathrm{D}$) are estimates of the radial velocities ($v_r$) (with an opposite sign, the Doppler velocities being positive towards the probe) corrupted by noise ($\eta$):

$$v_r(r, \theta, \varphi) = -u_\mathrm{D}(r, \theta, \varphi) + \eta(r, \theta, \varphi) \qquad (1)$$

The objective of the 3D-iVFM was to deduce the radial, polar, and azimuthal velocity components $[v_r(r, \theta, \varphi), v_\theta(r, \theta, \varphi)$ and $v_\varphi(r, \theta, \varphi)]$ in the full left ventricular volume, from the Doppler velocities ($u_D$) of the three planes. We extended the 2D-iVFM algorithm proposed by Assi *et al.* [12] and added physics-based equality constraints to obtain a hemodynamically valid estimate [25]. The hemodynamic equality constraints, based on fluid dynamics, ensured that the reconstructed flow was divergence-free (mass conservation for an incompressible fluid) and sliped freely on the endocardial wall (free-slip boundary conditions). We wrote the optimization problem as a constrained regularized least-squares problem, which was solved using the Lagrange multiplier method [26]. It was mathematically written as follows:

$$\vec{v}_{\text{VFM}} = \underset{\vec{v}}{\arg\min} J(\vec{v}) \qquad (2)$$

From (1), the cost function was written as

$$J(\vec{v}) = \int_\Omega (v_r + u_D)^2 \, d\Omega + \alpha \mathcal{L}(\vec{v}) \qquad (3)$$

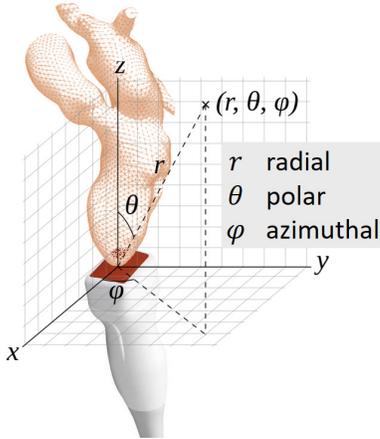

Figure 3 – The data and equations that are included in 3D-iVFM (intraventricular Vector Flow Mapping) are expressed in this spherical coordinate system.

The cost function $J$ was minimized subject to the following equality constraints:

$$\begin{cases} \text{div}(\vec{v}) = 0 & \text{on } \Omega, \\ \vec{v} \cdot \vec{n}|_{\text{wall}} - \vec{v}_{\text{wall}} \cdot \vec{n}|_{\text{wall}} = 0 & \text{on } \partial\Omega. \end{cases} \qquad (4)$$

$\vec{v}_{\text{VFM}}$ are the estimated three-component velocity vectors. The function $\mathcal{L}(\vec{v})$ induces spatial smoothing. It contains second-order partial derivatives of the velocity, with cross-terms, with respect to the radial and polar coordinates $(r, \theta)$. We used the same smoothing function as in Assi *et al.* [12]. The parameter $\alpha$ is the regularization smoothing parameter (a scalar), which was determined by using the $L$-curve method (see section "*Selection of the Smoothing Parameter*"). $\Omega$ stands for the intracavity region of interest, and $\partial\Omega$ is its boundary (endocardium). $\vec{n}|_{\text{wall}}$ is the unit vector normal to the endocardial wall, and $\vec{v}_{\text{wall}}$ is the endocardial wall velocity. The first equality constraint in (4) ensures mass conservation. By expressing the

divergence in spherical coordinates then multiplying by $r \sin(\theta)$, it can be rewritten as

$$\begin{aligned} \text{div}(\vec{v}) = 0 \implies \\ 2\sin(\theta)\, v_r + r\sin(\theta)\, \partial_r v_r + \\ \cos(\theta) v_\theta + \sin(\theta)\, \partial_\theta v_\theta + \partial_\varphi v_\varphi = 0 \text{ on } \Omega. \end{aligned} \qquad (5)$$

The second equality imposes free-slip boundary conditions. Because we work with 2-D images (planes), the azimuthal component of $\vec{n}|_{\text{wall}}$ ($n_\varphi|_{\text{wall}}$) is not available. To ensure well-posedness of the 3D-iVFM problem, we assumed that $v_\varphi = 0$ on $\partial\Omega$. The boundary conditions [second constraint in Eq. (4)] were thus rewritten as:

$$\begin{aligned} (v_r - v_{\text{wall } r})\, n_{\text{wall } r} + (v_\theta - v_{\text{wall } \theta})\, n_{\text{wall } \theta} = 0, \\ \text{and } v_\varphi = 0 \text{ on } \partial\Omega. \end{aligned} \qquad (6)$$

The spherical 3-D grid used for 3D-iVFM has constant radial, polar, and azimuthal steps ($h_r$, $h_\theta$, and $h_\varphi$). We used a finite difference discretization scheme (with second order central differences) to convert this constrained least-squares problem into a matrix-vector form. The Doppler velocities were stored in a 3rd-order tensor of size ($M \times N \times O$), where $M$ is the number of samples per scanline, $N$ is the number of scanlines per half-plane, and $O$ is the number of half-planes (i.e. $O = 6$ for a tri-plane mode, Figure 4). The numerical counterpart of the constrained least-squares problem described by Equations (2) to (6) was written using a tensor formalism, then vectorized (Figure 4), to obtain a sparse symmetric linear system. We briefly describe the linear system in the following.

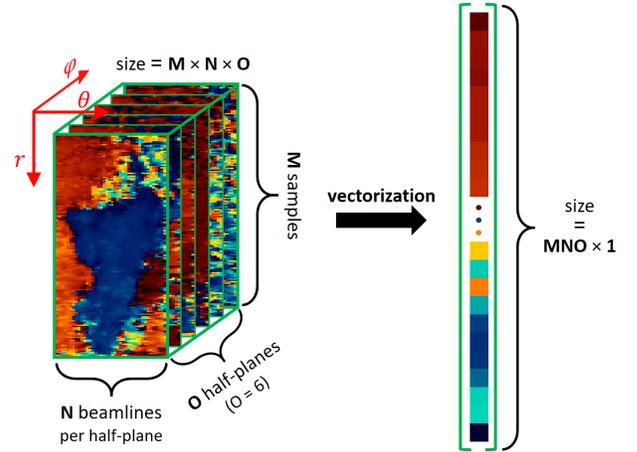

Figure 4 – The 3D-iVFM was derived using a 3rd-order tensor formalism. The equations were then vectorized to obtain a sparse symmetric linear system. The dimensions of the tensors were related to the coordinates $(r, \theta, \varphi)$ of the spherical system.

The vectorization led to column vectors. We note these column vectors as follows:

- $u_D$, of length ($MNO$), contains the Doppler velocities.
- $r, \theta, \varphi$, of length ($MNO$), contain the radial, polar, and azimuthal coordinates, respectively.
- $n_r$ and $n_\theta$, of length ($MNO$), contain the radial and polar coordinates of the unit vector normal to the cardiac inner wall

(endocardium). They are zeros if the node does not belong to the endocardium.

- $\boldsymbol{\delta}$, of length ($MNO$), is a binary vector that represents the intracavitary region of interest (ROI). It is 1 if the node is inside or on the edge of the left ventricular cavity, 0 otherwise.

- $\partial\boldsymbol{\delta}$, of length ($MNO$), is a binary vector that represents the endocardial wall. It is 1 if the node is on the endocardium, 0 otherwise.

- $\boldsymbol{v} = [\boldsymbol{v_r}^{\mathrm{T}} \ \boldsymbol{v_\theta}^{\mathrm{T}} \ \boldsymbol{v_\varphi}^{\mathrm{T}}]^{\mathrm{T}}$, of length (3 $MNO$), is a column vector that contains the iVFM solution of the minimization problem.

- $\boldsymbol{v}_{\mathrm{wall}} = [\boldsymbol{v}_{\mathrm{wall},r}^{\mathrm{T}} \ \boldsymbol{v}_{\mathrm{wall},\theta}^{\mathrm{T}} \ \boldsymbol{v}_{\mathrm{wall},\varphi}^{\mathrm{T}}]^{\mathrm{T}}$, with $\boldsymbol{v}_{\mathrm{wall},\varphi} = 0$ everywhere [see Eq. (6)], of length (3 $MNO$), is a column vector that contains the radial, polar and azimuthal (= 0) components of the endocardial velocities. They are zeros if the node does not belong to the endocardium.

- $\boldsymbol{\lambda_1}$ and $\boldsymbol{\lambda_2}$, of respective lengths ($MNO$) and (2 $MNO$), contain the Lagrangian multipliers associated with the 1st (divergence-free) and 2nd (boundary) equality constraints.

The iVFM minimization problem contains the $Q_0$, $Q_1$, $Q_2$, and $Q_\mathcal{L}$ matrices that are introduced in the next paragraph. The matrix $Q_0$ is related to the fitting to Doppler velocities (3). $Q_1$ and $Q_2$ are related to the 1st (5) and 2nd equality constraints (6). $Q_\mathcal{L}$ is associated with the smoothing regularizer included in (3). We also define the following matrices and vectors:

- $\dot{D}_q$ and $\ddot{D}_q$, both of size ($q \times q$), are finite-difference matrices for the 1st and 2nd derivatives, respectively.

- $\overline{\overline{0}}_q$ and $I_q$, both of size ($q \times q$), are the zero and identity matrices, respectively.

- $\boldsymbol{0}_q$ is vector of zeros of size ($q \times 1$).

In the following, $\otimes$ and $\circ$ stand for the Kronecker and Hadamard (element-wise) products. A least-squares numerical solution of the minimization problem given by (2) and (3), subject to the equality constraints (5) and (6), can be determined through the Lagrange multiplier method [26]. It is given by seeking the vectors $\boldsymbol{v}$, $\boldsymbol{\lambda_1}$, and $\boldsymbol{\lambda_2}$ that minimize

$$J_L(\boldsymbol{v}, \boldsymbol{\lambda_1}, \boldsymbol{\lambda_2}) = (Q_0\boldsymbol{v} + \boldsymbol{u_D})^{\mathrm{T}}(Q_0\boldsymbol{v} + \boldsymbol{u_D}) + \alpha(Q_\mathcal{L}\boldsymbol{v})^{\mathrm{T}}(Q_\mathcal{L}\boldsymbol{v}) + \boldsymbol{\lambda_1}^{\mathrm{T}}Q_1\boldsymbol{v} + \boldsymbol{\lambda_2}^{\mathrm{T}}Q_2\boldsymbol{v} \quad (7)$$

The $Q$-matrices are given by (supplementary document):

$$Q_0 = [1 \ 0 \ 0] \otimes \mathrm{diag}(\boldsymbol{\delta})$$

$$Q_1 = \left[ 2 \, \mathrm{diag}(\sin(\boldsymbol{\theta}) \circ \boldsymbol{\delta}) + \frac{1}{h_r} \mathrm{diag}(\boldsymbol{r} \circ \sin(\boldsymbol{\theta}) \circ \boldsymbol{\delta}) \, (I_O \otimes I_N \otimes \dot{D}_M), \right.$$
$$\mathrm{diag}(\cos(\boldsymbol{\theta}) \circ \boldsymbol{\delta})$$
$$+ \frac{1}{h_\theta} \mathrm{diag}(\sin(\boldsymbol{\theta}) \circ \boldsymbol{\delta}) \, (I_O \otimes \dot{D}_N \otimes I_M),$$
$$\left. \frac{1}{h_\varphi} \mathrm{diag}(\boldsymbol{\delta}) \, (\dot{D}_O \otimes I_N \otimes I_M) \right] \quad (8)$$

$$Q_2 = \begin{bmatrix} \mathrm{diag}(\boldsymbol{n_r}), & \mathrm{diag}(\boldsymbol{n_\theta}), & \overline{\overline{0}}_{MNO} \\ \overline{\overline{0}}_{MNO}, & \overline{\overline{0}}_{MNO}, & \mathrm{diag}(\partial\boldsymbol{\delta}) \end{bmatrix}$$

$$Q_\mathcal{L} = \begin{bmatrix} I_3 \otimes \left( \frac{1}{h_r^2} \mathrm{diag}(\boldsymbol{r} \circ \boldsymbol{r} \circ \boldsymbol{\delta}) \, (I_O \otimes I_N \otimes \ddot{D}_M) \right) \\ I_3 \otimes \left( \frac{1}{h_\theta^2} \mathrm{diag}(\boldsymbol{\delta}) \, (I_O \otimes \ddot{D}_N \otimes I_M) \right) \\ I_3 \otimes \left( \frac{\sqrt{2}}{h_r h_\theta} \mathrm{diag}(\boldsymbol{r} \circ \boldsymbol{\delta}) \, (I_O \otimes \dot{D}_N \otimes \dot{D}_M) \right) \end{bmatrix}.$$

The $Q_0$, $Q_1$, $Q_2$, and $Q_\mathcal{L}$ matrices are of size ($MNO \times 3MNO$), ($MNO \times 3MNO$), ($2MNO \times 3MNO$), and ($9MNO \times 3MNO$), respectively. The Kronecker products between the identity and $D$ matrices are related to the 1st- and 2nd-order derivatives. Minimizing $J_L(\boldsymbol{v}, \boldsymbol{\lambda_1}, \boldsymbol{\lambda_2})$ yields the linear system

$A\boldsymbol{x} = \boldsymbol{b}$, with

$$A = \begin{bmatrix} 2(Q_0^{\mathrm{T}}Q_0 + \alpha Q_\mathcal{L}^{\mathrm{T}}Q_\mathcal{L}) & Q_1^{\mathrm{T}} & Q_2^{\mathrm{T}} \\ Q_1 & & \\ Q_2 & & \overline{\overline{0}}_{3MNO} \end{bmatrix}$$

$$\boldsymbol{x} = \begin{bmatrix} \boldsymbol{v} \\ \boldsymbol{\lambda_1} \\ \boldsymbol{\lambda_2} \end{bmatrix}, \qquad \boldsymbol{b} = \begin{bmatrix} -2Q_0\boldsymbol{u_D} \\ \boldsymbol{0}_{MNO} \\ Q_2\boldsymbol{v}_{\mathrm{wall}} \end{bmatrix} \quad (9)$$

The matrix $A$ is sparse symmetric and of size ($6MNO \times 6MNO$). $\boldsymbol{b}$ and $\boldsymbol{x}$ are column vectors of size ($6MNO \times 1$). The linear system (9) illustrates the basic principles of 3D-iVFM. Rather than solving directly this system, we have completed it by parameterizing the velocity field as now explained.

### Parametrization of the Velocity Field

The 3D-iVFM technique uses a clinical triplane mode, which provides three long-axis planes (i.e. six azimuthal half-planes, Figures 2 and 4). Although Doppler sampling is appropriate in the radial and polar directions, it is limited in the azimuthal direction (6 samples). To obtain proper azimuthal derivatives and ensure a full-volume reconstruction, we expressed the velocity components by a parametrized function periodic in the azimuthal direction:

$$v_k = a0_k(r, \theta) + a1_k(r, \theta)\cos(\varphi) + a2_k(r, \theta)\cos(2\varphi) + a3_k(r, \theta)\cos(3\varphi) + a4_k(r, \theta)\sin(\varphi) + a5_k(r, \theta)\sin(2\varphi),$$
with $k \in \{r, \theta, \varphi\}$ $\quad (10)$

The parametrized $v_k$ contained six parameters since we had six half-planes. These coefficients ($aN_r$, $aN_\theta$, and $aN_\varphi$, with $N = 0 \dots 5$) were stored in a column vector $\boldsymbol{a}$ of length ($18MN$) (3 coordinates × 6 half-planes × $MN$ samples per half-plane). The linear system (9) was modified to include the coefficients of the parametrized velocities: all $Q$ matrices were multiplied by sparse matrices containing the coefficients $\{1, \cos(\varphi), \cos(2\varphi), \dots\}$ of the expression (10). After parametrization of the velocity field, the linear system (9) became

$$A\boldsymbol{x} = \boldsymbol{b}, \text{ with } \boldsymbol{x} = \begin{bmatrix} \boldsymbol{a} \\ \boldsymbol{\lambda_1} \\ \boldsymbol{\lambda_2} \end{bmatrix}. \quad (11)$$

The $A$ matrix, and the $\boldsymbol{x}$ and $\boldsymbol{b}$ vectors, obtained after parametrization, are described in the supplementary document. Solving the system (11) yields the coefficients $\boldsymbol{a}$ and, in turn, the full-volume three-component velocity fields. The $A$ matrix is rank-deficient because it contains columns and rows of zeros, as the region of interest and its boundary do not cover the entire domain. After having discarded the null rows and columns to make the matrix full-rank and positive-definite, we solved the sparse linear system (11) in MATLAB (MathWorks, latest release) by using an LDL decomposition.

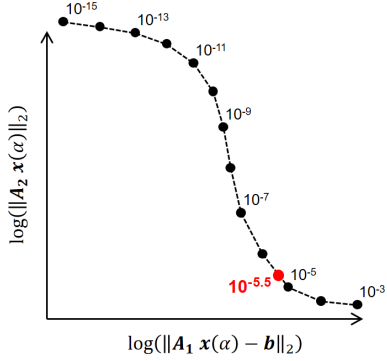

Figure 5 – Automatic selection of the regularization scalar: an L-curve is plotted from a set of regularization parameters (black dots). We choose the regularization parameter that provides the rising point of inflection (here $10^{-5.5}$, red dot). This example is from a patient at the end of early filling. In practice, we used five $\alpha$ values to generate an L-curve by fitting an hyperbolic tangent function.

### Selection of the Smoothing Parameter

The linear system (11) contains a regularization scalar ($\alpha$) that adjusts the effect of smoothing given by the $Q_{\mathcal{L}}$ matrix. To make the iVFM technique unsupervised, we selected the smoothing parameter automatically through the L-curve method [27]. This method helps in identifying the trade-off between the amount of regularization (smoothing) and the quality of the fit to the input (Doppler) data. The L-curve consists of a log-log plot of the residual norm versus the regularization norm for a set of $\alpha$ values. In our case, the "optimal" $\alpha$ was the one that corresponded to the rising point of inflection of the L-curve. Figure 5 depicts an L-curve obtained in one patient (see also Figure 10). To obtain the L-curve, we decomposed the matrix $A$ into two parts: a constrained data-fitting matrix $A_1$, and a matrix $A_2$ related to the smoothing, which reads as follows:

$A = A_1 + \alpha A_2$, where

$$A_1 = \begin{bmatrix} 2Q_0^{\mathrm{T}}Q_0 & Q_1^{\mathrm{T}} & Q_2^{\mathrm{T}} \\ Q_1 & \bar{\bar{0}}_{3MNO} \\ Q_2 \end{bmatrix} \tag{12}$$

$$A_2 = \begin{bmatrix} 2Q_{\mathcal{L}}^{\mathrm{T}}Q_{\mathcal{L}} & \bar{\bar{0}}_{3MNO} \\ \bar{\bar{0}}_{3MNO} & \bar{\bar{0}}_{3MNO} \end{bmatrix}$$

From these matrices, the L-curve associated with the linear system (11) was given by

$$\{\log(\|A_1\,\boldsymbol{x}(\alpha) - \boldsymbol{b}\|_2)\,,\log(\|A_2\,\boldsymbol{x}(\alpha)\|_2)\}. \tag{13}$$

The L-curve method requires solving the linear system with several values of $\alpha$. We used five $\alpha$ parameters to generate an

L-curve and determine its "optimal" $\alpha_{opt}$ value. To reduce computational time, it is preferred not to repeat this process for each triplane color-Doppler. Therefore, we calculated the L-curve, and its corresponding optimal $\alpha_{opt}$, once, at the end of early filling. We then used this same $\alpha_{opt}$ value for the whole cardiac cycle.

### In Silico Validation

Validating the 3D-iVFM algorithm required full-volume three-component ground-truth velocity fields of the intraventricular blood flow. We used a patient-specific CFD (computational fluid dynamics) model [28], [29], as for the latest version of 2D-iVFM [12]. We interpolated the radial components (Figure 3) of the CFD velocity vectors onto three planes (as in Figure 2) to simulate triplane color Doppler. One hundred triplane echocardiographic scans were simulated for one cardiac cycle. The radial and polar steps were 0.55 mm and 0.45°. Each half-plane contained 8,000 Doppler-velocity samples ($M = 160$, and $N = 50$). Gaussian white noise with velocity-dependent local variance (signal-to-noise ratio ranging between 20 and 50 dB with a 10-dB step) was added, as in [30]. Based on these simulated triplane Doppler velocities, the coefficients of the parametrized iVFM velocities, given by (10), were calculated by solving the linear system (11). From these coefficients, we obtained full-volume three-component intraventricular velocity fields on an $M \times N \times 24$ spherical grid (i.e. on 24 half-planes). We compared the iVFM-derived velocity components against the actual CFD-based velocity components by regression analysis and by calculating the normalized root-mean-square errors (nRMSE) for each noise level:

$$\text{nRMSE} = \frac{1}{\max(\|\bar{\boldsymbol{v}}_{\text{CFD}}\|)} \sqrt{\frac{1}{LVV}\iiint \|\boldsymbol{v}_{k-\text{CFD}} - \boldsymbol{v}_{k-\text{iVFM}}\|^2\,r^2\sin(\theta)\;\mathrm{d}r\mathrm{d}\theta\mathrm{d}\varphi} \tag{14}$$

with $k \in \{r, \theta, \varphi\}$.

In Eq. (14), $LVV$ stands for the left ventricular volume. In addition, we computed the radial, polar, and azimuthal components of the vorticity, as well as its amplitude, for both CFD and iVFM, and we compared their means. We finally estimated the vortex volumes by using the $Q$-criterion [31]: a voxel belonged to a vortex if its $Q$-value was greater than a predefined threshold. The threshold was ranged between 5,000 and 15,000 s⁻² to get several volume estimates per time step. We compared the mean vortex volumes (CFD vs. iVFM) over diastole. The vorticities ($\omega$), $Q$-criterion, and vortex volumes of the CFD flow data were calculated using ParaView (Kitware Inc., NY, USA), an open-source package for data analysis and visualization. Those derived by iVFM were estimated through in-house MATLAB codes.

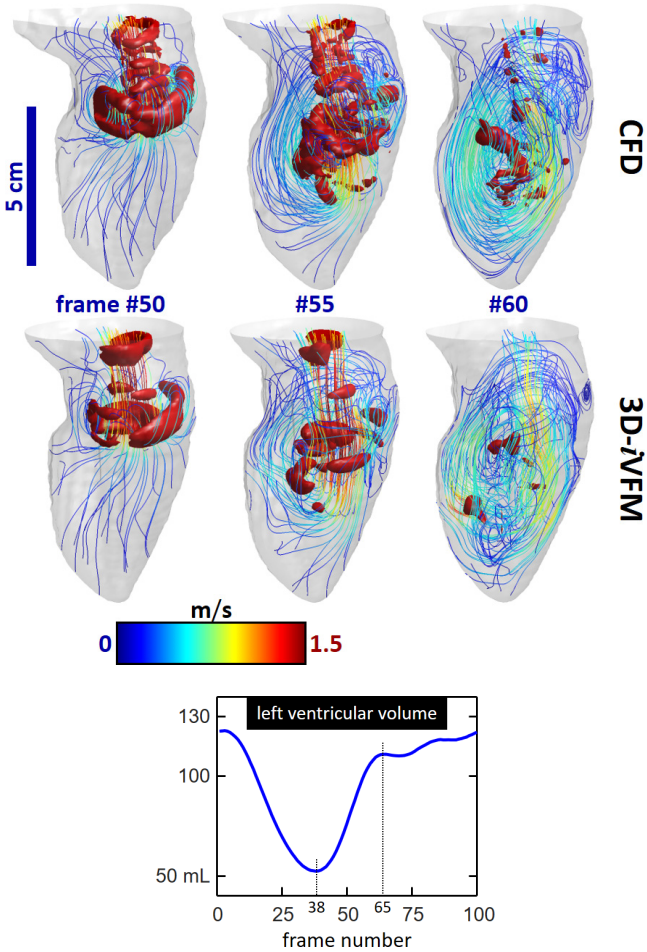

Figure 6 – *In silico* results – 3D-iVFM (2ⁿᵈ row) enables the reconstruction of the intraventricular blood flow. The 1ˢᵗ row shows the original CFD velocity fields. The colors of the streamlines represent the velocity amplitudes. The red volumes illustrate the vortex regions detected by the $Q$-criterion. In the CFD model, the left ventricular relaxation started at frame #38 and ended at frame #65.

### In Vivo Analyses

The clinical feasibility of 3D-iVFM was tested in four volunteers, with or without cardiac disease, during a routine echocardiographic examination. Triplane echocardiography is part of the standard examination protocol. The retrospective study of anonymized data was approved by the local ethics committee (Hospices Civils de Lyon, approval #69HCL20-0774). The cardiologist acquired the two-, three-, and four-chamber apical long-axis views using a triplane color Doppler mode with a GE Vivid E95 scanner (Figure 1). We selected triplane echocardiographic data with good-quality B-mode and Doppler images (echogenic patients, complete left ventricle in all three planes, no visible artifacts). We obtained ~11 color-Doppler triplanes per cardiac cycle. The Doppler velocities and B-mode images (before scan-conversion, Figure 2) were saved in Hierarchical Data Format with the EchoPAC clinical software package (GE Healthcare) then post-processed using MATLAB programming. We delineated the endocardial walls manually by using an in-house MATLAB application. The wall velocities were estimated from the endocardial displacements between two successive frames. We finally computed the full-volume three-component intraventricular velocity fields with the 3D-iVFM method. Creating the sparse matrices and solving the linear system required ~20 s with a MATLAB code running on the CPU. The vortical regions were detected by using the $Q$-criterion, and the intraventricular blood flows were visualized through streamlines.

### RESULTS

#### In Silico Results

As illustrated by Figure 6, 3D-iVFM enabled the reconstruction of full-volume three-component intraventricular velocity fields from three color-Doppler long-axis planes (triplane color Doppler). In particular, the vortex rings that formed during left ventricular filling (frame #50), and the large rotating circulation at the end of filling (frame #60) were recovered. The iVFM-derived flow structures were consistent with those of the CFD ground-truth.

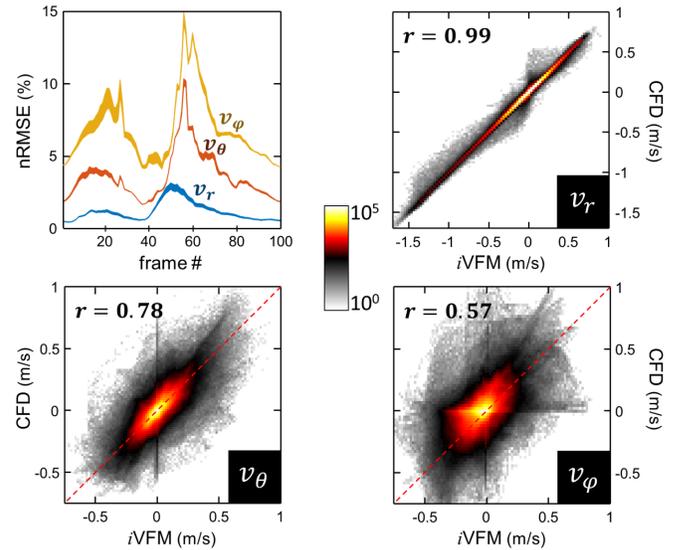

Figure 7 – *In silico* results: velocities derived by 3D-iVFM vs. actual CFD velocities. Top-left figure: normalized root-mean-square errors (nRMSE) on the radial, polar, and azimuthal velocity components. The thickness of the curves reflects the range due to the signal-to-noise ratio (20 to 50 dB). Other figures: iVFM vs. CFD velocities for each component after pooling all the frames (SNR = 30 dB). The colors represent the number of voxels that fall in each bin of the bivariate histogram. The dashed red line is the identity line. The $r$ are the correlation coefficients.

The velocity errors were the smallest for the radial $r$-components (less than 3%), while the errors on the azimuthal $\varphi$-components ranged between 5% and 15% (Figure 7). The regression analyses confirmed these results. The 3D-iVFM technique reconstructed the radial components with accuracy ($r = 0.99$) since the input information from color Doppler is essentially radial. The regression coefficients obtained with the polar and azimuthal components were $r = 0.78$ and $r = 0.57$, respectively.

The errors on the radial, polar, and azimuthal components reflect the voxelwise (local) biases in velocity estimation. Figure 8 and Figure 9 show how global vorticity parameters evolved. The volumes of the vortex cores (Figure 8), as determined by the $Q$-criterion (red volumes in Figure 6), were the largest at mid-filling (around frame #55). The vortex volumes derived from iVFM agreed with those measured in the CFD model, but contained some inconsistencies (Figure 8). The peak of the mean vorticity amplitude was reached (80 to 90 s$^{-1}$) around frame #55 (Figure 6, and Figure 9, left). The azimuthal vorticity components ($\omega_\varphi$) were the most prominent (Figure 9, right) compared with the radial ($\omega_r$) and polar ($\omega_\theta$) components. 3D-iVFM produced vorticity profiles consistent with those given by CFD (ground-truth).

Figure 11 displays diastolic flow streamlines in three patients with heart disease.

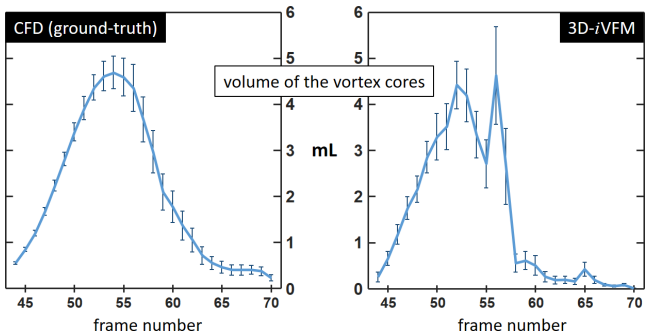

Figure 8 – *In silico* results: vortex volumes derived by 3D-iVFM vs. actual CFD velocities. The vortical regions were detected by using the $Q$-criterion. A voxel belonged to a vortex if its $Q$-value was greater than a threshold. Several thresholds were used: the error bars illustrate the standard errors of the means (SEM).

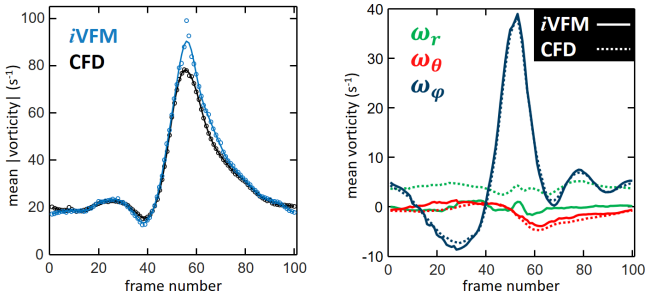

Figure 9 – *In silico* results: mean vorticities derived by 3D-iVFM vs. actual CFD velocities. Left panel – mean of the vorticity amplitudes. Right panel – mean of the vorticity components: radial ($\omega_r$), polar ($\omega_\theta$), azimuthal ($\omega_\varphi$).

### 3D-iVFM in Patients

Our 3D-iVFM technique enabled the reconstruction of intraventricular flows in selected patients after a routine echocardiographic examination. Figure 10 displays the flow streamlines in a normal patient (no cardiac disease) at different stages of the cardiac cycle. The vortex ring formed by the mitral jet at the onset of early filling (i.e., ventricular relaxation) is visible. A large long-axis circular motion then occupies the entire intraventricular cavity at the end of early filling and during diastasis (period between ventricular relaxation and atrial contraction).

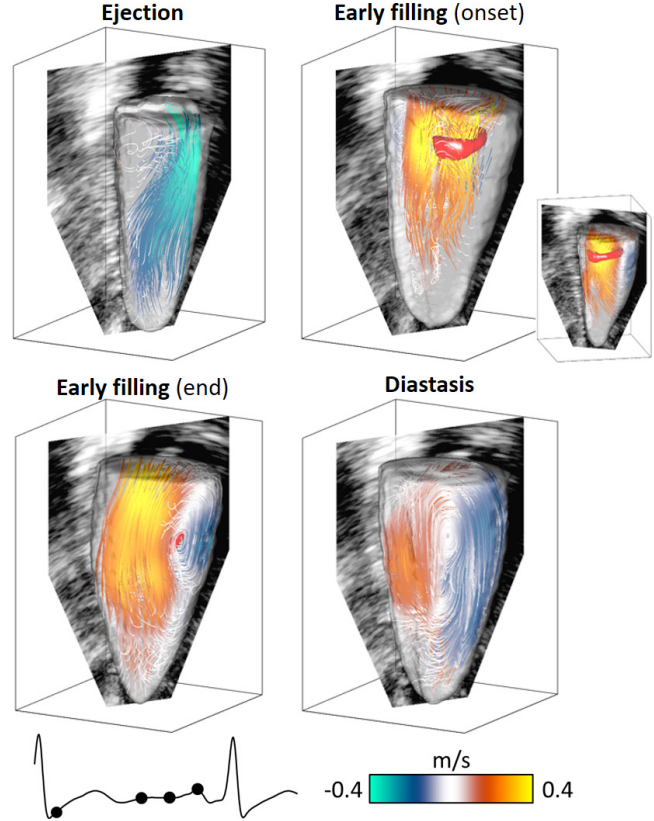

Figure 10 – 3D-iVFM in a normal patient. The colors represent the Doppler velocities. The vortex ring is visible at the onset of early filling (left ventricular relaxation).

## Discussion

### Quality of the Reconstructed Vector Field

The 3D-iVFM introduced in this article allows reconstruction of full-volume three-component intraventricular flows from triplane color Doppler. This mode provides three long-axis apical views separated by an azimuthal angle of sixty degrees. In the spherical coordinate system associated with the cardiac probe, the triplane mode measures the radial components of the velocities, with noise. The 3D-iVFM problem has essentially two unknowns, which are the polar and azimuthal angular components of the velocities. For comparison, only the polar components are sought with the 2D-iVFM that operates in a three-chamber view. As we did for the 2D-iVFM, the 3D-iVFM was validated using an intracardiac CFD model. This CFD model has the advantage of providing very realistic reference intraventricular flows at high spatial and temporal resolutions.

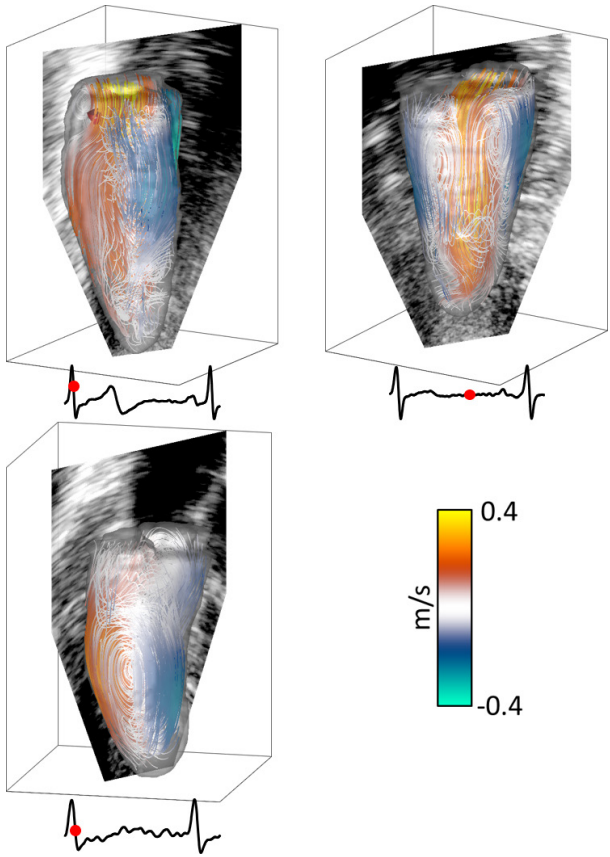

Figure 11 – 3D-iVFM in three patients. The colors represent the Doppler velocities. Left column: a vortex is present during the isovolumetric contraction. Top-right: a vortex ring is visible at the onset of atrial systole.

As expected, the radial components of the velocity field were the best reconstructed, with errors less than 3%. The differences originated from the smoothing and inter-angle interpolation imposed by the iVFM. The errors on the polar components were close to those observed with the recent version of the 2D-iVFM [25], ranging between 2 and 10%. On the other hand, the azimuthal components were the most error-prone because we had little information in this direction. Only six samples (six half-planes) were available. To compensate for this azimuthal undersampling, we used a parametric periodic formulation of the velocities, given by Equation (10). The errors in velocities were greatest at the beginning of ventricular filling. This is a phase in the flow where polar and azimuthal velocities become significant. The vortex changes from a ring to a recirculation shape. Despite significant azimuthal errors, the global aspect of the flow was adequately recovered (Figure 9). More importantly, the global vorticity was determined accurately, which tends to show that the 3D-iVFM could be used for physiological purposes to characterize the intracavitary left ventricular flows.

### Technical and Numerical Limitations

Intracardiac blood flow is three-dimensional and changes rapidly during a cardiac cycle. A conventional triplane color Doppler returns around 10 triplane sequences per second. To obtain a triplane sequence, a series of focused or narrow ultrasound beams are transmitted sequentially in the radial direction to scan the planes of interest. The spatial resolution of one plane is approximately 0.6 mm by 1.4 degrees. In addition, for each radial direction, about six ultrasound beams are transmitted to enable Doppler velocities to be determined. It results that there is a significant time delay between the first Doppler line of the first plane and the last Doppler line of the third plane. A triplane dataset, therefore, has a limited spatiotemporal resolution. Given the limitations of triplane color Doppler, 3D-iVFM can only provide smoothed estimates of the velocity fields. Although several researchers have reported measuring energy dissipation due to blood viscosity in turbulent flow by two-dimensional iVFM [32], [33], this claim is incorrect. The fluctuating quantities of turbulent flow cannot be captured by color Doppler. Nevertheless, the results obtained with the CFD model show that it would be possible to measure global vortical indices (Figure 9). In particular, we anticipate that the peak of global vorticity could reflect the filling function. This remains to be demonstrated, as our study does not allow such a conclusion. We qualitatively tested the algorithmic methodology, and evaluated its feasibility in a clinical setting. For this purpose, we generated noisy "ideal" Doppler fields from the radial velocity components given by the CFD model. Our results thus do not allow us to conclude on the effects of ultrasound imaging artifacts, such as clutters related to wall motions and sidelobes or grating lobes, and dropouts due to clutter filtering. A study coupling fluid dynamics and acoustics would complete the validations and provide insights into these issues. An approach would be the use of ultrasound simulations [34], [35] after seeding the CFD flow with scattering particles [36], [37].

### From 2-D, to 3-D, to 4-D?

The iVFM that we introduced in this work is a three-dimensional upgrade of the previous 2-D version [25]. We could even claim that it is a 4-D version (volume + time) because temporal information is also available. The temporal resolution (~10 volumes/s) is nevertheless limited by that of triplane color Doppler. The rate of iVFM volumes is similar to that of 4-D MRI velocimetry. Under the condition that the heart rate is sufficiently stable, it would then be necessary to analyze more than five cardiac cycles to decipher the flow dynamics with sufficient accuracy. This is precisely the approach used in 4-D flow MRI, with respiratory navigator gating. The temporal limitation of triplane echocardiography is related to the sectorial line-by-line scanning of each of the three planes. A promising solution would be the use of diverging rather than focused waves [38], [39]. Such an alternative of high-frame-rate triplane echocardiography, with simulated spiral arrays, has been proposed by Ramalli *et al.* using multi-line transmits and diverging waves [40]. Another approach would be the use of high-volume-rate color Doppler [41], which would have the twofold advantage of increasing temporal resolution and providing more Doppler information (full volumes instead of three planes). We have not tested these approaches because they are not yet available for clinical examinations.

### What Still Needs to be Done

We showed that three-dimensional reconstruction of intracardiac flow is feasible during a routine clinical examination. We used triplane echocardiography and Doppler data before scan-conversion through GE Healthcare EchoPAC clinical software. The time for ultrasound acquisition (<20 s) and 3-D velocity calculation (<20 s per volume), was well below that of 4-D flow MRI, even though we used a homemade MATLAB code. As discussed previously, volumetric Doppler data would likely allow more accurate 3-D flow reconstruction, albeit at the expense of temporal resolution. These volume data are generally protected and therefore not accessible for post-processing. Grønli *et al.* (in conjunction with GE Healthcare) showed in a conference paper [20] that iVFM with volume data is possible. Using TensorFlow, they regularized the velocity fields by imposing mechanical constraints and endocardial boundary conditions. They also succeeded in increasing the volume rate by transmitting wide ultrasound waves. Rather than Doppler velocities, blood speckle tracking has also been proposed for the 2-D reconstruction of intracardiac flow [10], [14], [15]. It is possible to combine color Doppler with speckle tracking to get the most out of each. This has been achieved for 2-D myocardial motion analysis using diverging-wave echocardiography [42]. This approach could theoretically be applicable in 3-D for intracardiac flows. In addition to beamforming, motion detection, and regularization methods, iVFM is also dependent on clutter filtering, dealiasing, and endocardial segmentation. Having made use of a clinical ultrasound device, we did not have control over the wall filter. The dealiasing method that we used was an unwrapping technique introduced by Muth *et al.* [30]. It depends on an input variable that sometimes had to be adjusted manually. To make dealiasing fully automatic, we will resort to deep learning [43]. In this study, the endocardial segmentation was performed manually for the analysis of the clinical cases. This step was necessary to determine the boundary conditions. To avoid this time-consuming task, a future version of 3D-iVFM will include deep learning-assisted segmentation, as described in Leclerc *et al.* [44]. Although there is some room for technical improvements, the 3D-iVFM version introduced in our study shows that it is possible to visualize, and quantify, left intraventricular flows in three dimensions. The 3D-iVFM could be applicable in a clinical setting since it uses a standard apical echocardiographic window and fast algorithms.

## CONCLUSION

We have introduced an improved Doppler-based iVFM algorithm for recovering three-component full-volume intraventricular blood flows. For this purpose, we used a minimization problem constrained by equations consistent with fluid dynamics. The model selection was automatized through an *L*-curve to minimize operator-dependence. The 3D-iVFM is compatible with standard echocardiographic examination and is not time-consuming.